# Nonequilibrium Work Fluctuations in Force-induced Melting of a Short B-DNA

S. Siva Nasarayya Chari[1,*] and Prabal K. Maiti[2]

[1]*Department of Physics, Faculty of Science, University of Allahabad, Prayagraj – 211002, Uttar Pradesh, India.*
[2]*Centre for Condensed Matter Theory, Department of Physics, Indian Institute of Science, C V Raman Ave, Bengaluru – 560012, Karnataka, India.*

[*]Corresponding author: snchari@allduniv.ac.in

**Abstract.** A system of a solvated canonical B-DNA of 12-base pairs with the specified sequence is initially equilibrated in a state of zero external force $f$ acting on it. After equilibration, a switching experiment is performed over the system by pulling one end of the DNA while restraining its other end. The finite time pulling process is performed at a constant rate of the applied force, until a maximum value of $400\ pN$. The associated nonequilibrium work done ($W$) during this process is determined by numerically integrating the force-extension curve, as a function of the applied force. An ensemble of the work values, $p(W)$ is obtained by repeating the pulling experiments a large number of times. The free energy difference ($\Delta F$) between the equilibrium and force-induced melted states of the DNA is determined by employing the Jarzynski equality. The value of $\Delta F$ is found to be in close agreement with the conventional equilibrium methods.

## INTRODUCTION

Understanding the processes of DNA denaturation, replication of DNA, and their associated free energy differences is important to gain insight into fundamental mechanisms that happen at the level of nucleosomes. Free energy calculations of the molecules like DNA, and RNA under various physiological conditions are highly useful in drug manufacturing. A short double-stranded DNA (dsDNA) can be made to undergo denaturation either by heating or by applying an external force. The latter method is referred to as the force-induced melting [1] of dsDNA. This is a finite-time process in which DNA is switched from an initial equilibrium state $A$ to a final nonequilibrium, unzipped state $B$. If the pulling process is not so far from equilibrium, then one can invoke the Jarzynski equality(JE) [2] or the Crooks Fluctuation Theorem(CFT) [3] to estimate the equilibrium free energy difference $\Delta F$, between the two states of the DNA. Here, we consider a 12-base pair canonical B-DNA and determine its equilibrium free energy difference along the pathway of force-induced melting, by employing the JE. Methods of the Simulation and the results are discussed in the following section.

## NUMERICAL EXPERIMENT

The structure of a 12-base pair double-stranded B-DNA with the sequence *CGCGAATTCGCG* is created with the LEaP module of the AMBER 10 software [4]. The *parambsc0* force field parameters are used to describe the molecular interactions associated with the DNA. $Na^+$ counter-ions (22 in number) are added with Joung-Cheatham parameters [4] to neutralize the charge on DNA. It is then solvated with the TIP3P [5] water molecules. It is made sure that the DNA is surrounded by about 10Å of solvent shell even in the fully stretched configuration. The system is then equilibrated at the temperature of 300 K with a standard equilibration protocol [1] and then an ensemble of equilibrium configurations of the system is collected to perform the switching experiment. A time-dependent pulling force,

$$f(t) = f_0 + (f_1 - f_0)\frac{t}{\tau} = f_0 + t(\delta f_\tau) \qquad (1)$$

is applied to one end of the DNA, on the $O3'$ and $O5'$ atoms in particular, while keeping the other end of the DNA fixed by applying a strong positional restraint. In eq. (1), the parameters $f_0 = 0$, $f_1 = 400\ pN$, and the rate of pulling, $\delta f_\tau = 10^{-4} pN/fs$ which is about two orders of magnitude faster than the laboratory experiment. The resulting extension of the DNA with the applied pulling force is depicted in Fig. 1(a). From the figure, we observe a structural transition in DNA around $150\ pN$ [8]. The number of Watson-Crick Hydrogen bonds (H-bonds) [10] that are intact between the base pairs of the DNA during the pulling process is shown in Fig. 1(b). The figure indicates that the maximum number of H-bonds was intact until about $150\ pN$. That means the original B-DNA has now transited to an S-DNA [6]. Further increasing $f(t)$ destroys the H-bonds, as evident from Fig. 1(b).

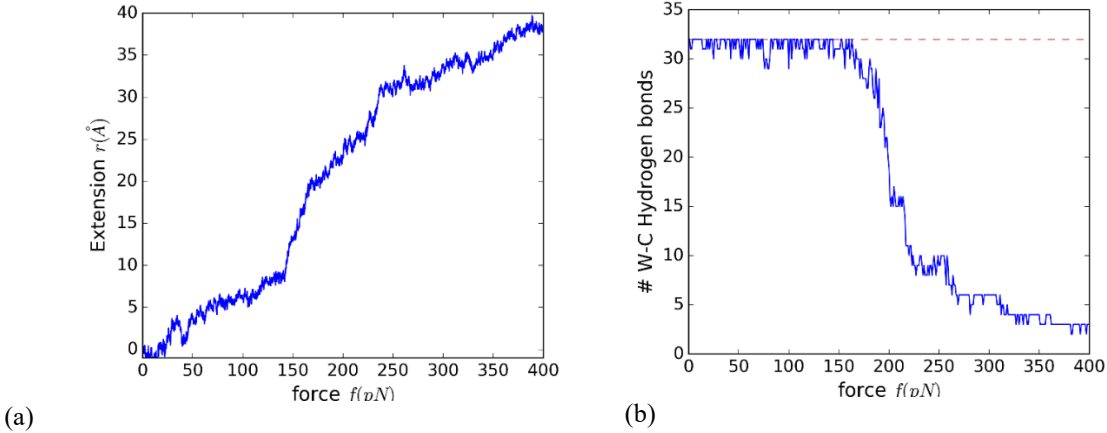

(a)         (b)

**FIGURE 1.** (a). Extension of the DNA molecule in Å observed with the applied pulling force in pico Newtons. (b). The number of Watson-Crick Hydrogen bonds that are intact in DNA as a function of the pulling force. The dashed line indicates the maximum possible number of 32 H-bonds.

The associated work done during this process is obtained as,

$$W = \int_0^{400\ pN} x(f)df, \qquad (2)$$

where $x(f)$ corresponds to the extension at force $f$. The above integral is numerically estimated using a trapezoidal rule. This constitutes one switching experiment. A hundred such pulling experiments were carried out, with each one starting from an initial equilibrium state A, of $f = 0$, and ending with the state B, of $f = 400\ pN$, may not be in equilibrium, yielding an ensemble of work values. Figure 2(a) shows the fluctuating work values with each performed experiment, at the chosen rate of pulling and the corresponding normalized work distribution $p(W)$ is presented in Fig. 2(b). From Fig. 2(b) we note that $\langle W \rangle > \Delta F$, implying a positive dissipation, $W_d = (\langle W \rangle - \Delta F) > 0$, satisfying the Callen-Welton theorem [7], and consistent with the Second law of Thermodynamics. However a small fraction of the work values in the left tail of $p(W)$, where $W < \Delta F$, would violate the Second law [8].

The equilibrium free energy difference $\Delta F$ between the states, $A$ and $B$ of the system could be estimated from the nonequilibrium work distribution $p(W)$ by employing the Jarzynski equality,

$$\left\langle e^{-\beta W} \right\rangle = e^{-\beta \Delta F},$$
$$\Rightarrow \Delta F = -\frac{1}{\beta} \ln \left\langle e^{-\beta W} \right\rangle, \qquad (3)$$

where $1/\beta = k_B T$, and the averaging $\langle ... \rangle$ is done over the ensemble of pulling trajectories. From eq. (3), we obtained $\Delta F = 91.2\ kcal/mol$, which is in close agreement with the equilibrium simulations performed [9] on the same system using the conventional Umbrella Sampling(US) method. However, the control parameter in methods like US is the molecule's end-to-end length $\xi$, unlike in a pulling process where force $f$ is controlled. Therefore, the trajectory of microstates the system goes through during the process, hence the unzipping pathways could be different for both methods. Despite this, the $\Delta F$ value agrees closely for both of them.

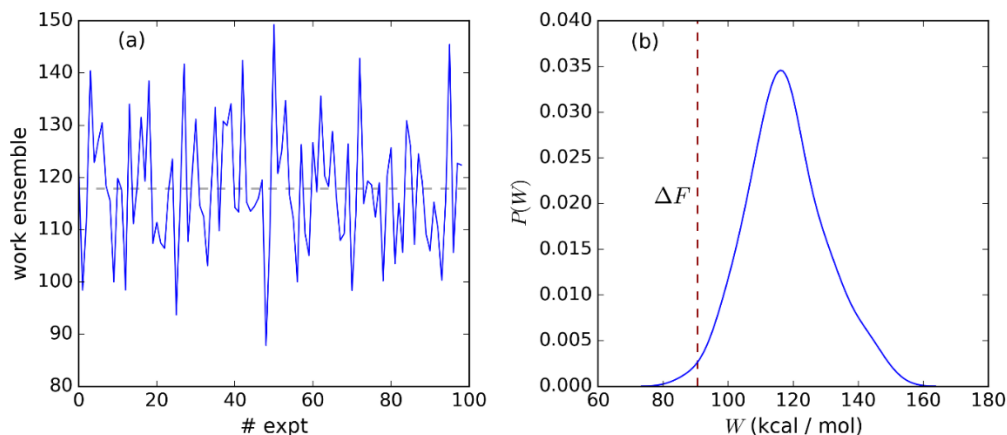

**FIGURE 2.** (a). Work values (in units of kcal/mol) with the corresponding experiment number; the dashed line represents the average work $\langle W \rangle$. (b). Normalized work distribution $p(W)$, the dashed line corresponds to the equilibrium $\Delta F$ (= 91.2 kcal/mol) obtained from the Jarzynski equality.

To summarize, a finite time switching process is performed on a canonical B-DNA with 12 base pairs. The equilibrium free energy difference between the natural zipped and the unzipped states of the DNA is determined using the Jarzynski equality, which is found to be in close agreement with the conventional Umbrella sampling method despite the possible differences in the sampled trajectory of microstates, and the pathway of unzipping. Therefore, Jarzynski equality is a viable nonequilibrium method for this system for free energy calculations. However, the work fluctuations become observable only for small systems. The average work $\langle W \rangle \to \Delta F$, and $p(W) \to \delta(W - \Delta F)$ as one tends to the thermodynamic limit.

## ACKNOWLEDGMENTS

SSNC thanks the Computational facility provided by the Department of Physics, Indian Institute of Science. The computational facility at the Solid State and Structural Chemistry Unit(SSCU), Indian Institute of Science, Bengaluru is highly acknowledged.